\documentclass[prb,aps,twocolumn,floats,floatfix]{revtex4}
\usepackage{graphicx}
\usepackage{amsmath}
\usepackage{amssymb}
\usepackage{amsfonts}
\def\nin{\noindent}

\def\be{\begin{equation}}
\def\ee{\end{equation}}
\def\bea{\begin{eqnarray}}
\def\eea{\end{eqnarray}}
\def\d{{\rm d}}
\def\r{{\bf r}}

\def\q{{\bf q}}

\def\n{{\bf n}}
\def\G{{\cal G}}
\def\D{{\cal D}}

\def\lp{\left(}
\def\rp{\right)}
\def\lb{\left[}
\def\rb{\right]}
\def\la{\left<}
\def\ra{\right>}

\def\Gbare{G^{(0)}}

\def\V{\widehat V}

\def\w{\omega}

\begin{document}
\title{Random walk with barriers:\\ Diffusion restricted by permeable membranes}

\author{Dmitry S. Novikov}
\affiliation{Center for Biomedical Imaging, Department of Radiology,
New York University School of Medicine, New York, NY 10016, USA}
\author{Els Fieremans}
\affiliation{Center for Biomedical Imaging, Department of Radiology,
New York University School of Medicine, New York, NY 10016, USA}
\author{Jens H. Jensen}
\affiliation{Center for Biomedical Imaging, Department of Radiology,
New York University School of Medicine, New York, NY 10016, USA}
\author{Joseph A. Helpern}
\affiliation{Center for Biomedical Imaging, Department of Radiology,
New York University School of Medicine, New York, NY 10016, USA}

\date{\today}

\begin{abstract}
\nin
Restrictions to molecular motion by barriers (membranes) are ubiquitous in biological tissues, porous media and composite materials.
A major challenge is to characterize the microstructure of a material or an organism nondestructively using a bulk transport measurement. Here we demonstrate how the long-range structural correlations introduced by permeable membranes give rise to distinct features of transport.
We consider Brownian motion restricted by randomly placed and oriented permeable membranes and focus on the disorder-averaged diffusion propagator using a scattering approach.
The renormalization group solution reveals a scaling behavior
of the diffusion coefficient for large times,
with a characteristically slow inverse square root time dependence.
The predicted time dependence of the diffusion coefficient agrees well with Monte Carlo simulations in two dimensions.
Our results can be used to identify permeable membranes as restrictions
to transport in disordered materials and in biological tissues, and to quantify
their permeability and surface area.
\end{abstract}

\maketitle

Brownian motion in a uniform medium is characterized by a single parameter,
the diffusion coefficent $D$,
which is a measure of mean square molecular displacement.
A packet of random walkers spreads with time $t$ according to a
Gaussian distribution with variance $\la x^2\ra = 2Dt$.
Complexity in the microscopic structure of a sample,
such as heterogeneity in diffusive properties and restrictions to molecular motion,
results in non-Gaussian evolution\cite{haus-kehr,Bouchaud}.
In particular, the diffusion coefficient $D(t)$ itself becomes time dependent. 

A fundamental question is how specific complexity features
manifest themselves in the coarse-grained dispersive dynamics.
Random drifts are known to drastically slow the dynamics down to
$\la x^2\ra\sim \ln^4 t$, thereby suppressing the
diffusion, $D|_{t\to\infty}=0$, in one dimension\cite{Sinai,Marinari}.
Their effect in higher dimensions is less profound
\cite{Fisher,Aronovitz-Nelson,Fisher-Shenker,KLY,Lerner94}.
Randomness in local diffusion coefficient preserves Gaussian diffusion at $t\to\infty$, resulting in the finite limit $D_\infty\equiv D(t)|_{t=\infty}$. However, it causes a power-law dispersion $\sim \w^{d/2}$ in the real part of the velocity autocorrelation function $\D(\w)=\int_0^\infty\! \d t\, e^{i\w t}\langle v(t)v(0)\rangle $; this power law depends on the spatial dimensionality $d$ and results in long time tails in the system's current-density response kernel $\D(t)$.\cite{Ernst-I}
Studying dispersive diffusion, therefore, is a way to characterize the type of disorder and of restrictions to molecular motion in a complex sample.

Here we consider the important class of restrictions to diffusion, the
permeable barriers (membranes).
Practically, membranes play an essential role for the transport of ions \cite{sykova-nicholson-2008},
water molecules \cite{cory-garroway,callaghan,latour-pnas,LeBihan},
and gases \cite{yablonskiy-lung-pnas} in biological tissues\cite{Friedman},
and in porous and composite materials\cite{cotts-nature1991,callaghan-nature1991,Mitra92,mair-prl1999,sen-nature,sen-perm}.
On a fundamental level, while occupying vanishing volume,
permeable membranes introduce important long-range correlations into the structure of a disordered sample. We show that these correlations give rise to distinct transport features, qualitatively different from those in ``uncorrelated" disordered systems\cite{Ernst-I}.

We develop a minimal model which appeals to any $d$-dimensional complex medium where the restrictions by permeable membranes play a dominant role.
We assume that diffusion is restricted by randomly placed and oriented infinite flat membranes, such as in the $d=2$ example of Fig.~1.
We find the diffusion coefficient as a function of time or frequency, for all membrane concentrations and permeabilities.
We demonstrate that the {spatially correlated disorder} introduced by the membranes results in a long-term memory that manifests itself in a distinct non-analytic low-frequency dispersion of the diffusion, $\D(\w)-D_\infty \sim \sqrt{\w}$. The latter is equivalent to a characteristically {slow} decrease of the diffusion coefficient, $D(t)-D_\infty \sim 1/\sqrt{t}$ as $t\to\infty$, Fig.~2, causing the mean square molecular displacement to increase as $\la x^2\ra \simeq 2D_\infty t + \mbox{const}\cdot \sqrt{t}$.
We relate this power law dispersion to the anomalously strong fluctuations of the amount of restrictions in a given volume, caused by the spatially extended nature of the disorder.
This makes
the $\sqrt{\w}\sim 1/\sqrt{t}$ dependence present in any spatial dimension $d$ as long as the membranes are flat on the scale of the diffusion length, as confirmed numerically for $d=2$.


Our finding emphasizes the role of spatial correlations in disordered samples in contrast to the short-range disorder typically considered in classical\cite{Ernst-I} and quantum\cite{Altshuler-Aronov} transport. It constitutes a novel disorder universality class of classical random media.
%
%
The $\sqrt{\w} \sim 1/\sqrt{t}$ dependence in the system's low-frequency dynamics can serve as a unique ``fingerprint" of the permeable membranes within the complexity of realistic samples in which the transport can be studied via
the electrical\cite{Scher73} or heat conduction, or by the diffusion-weighted NMR\cite{callaghan}.
%
%


\section*{Model}


We begin by outlining a minimal model of a sample in which the dominant restrictions to molecular motion are random permeable membranes.
A membrane is an idealization of a thin slice of a poorly diffusive material, as long as its thickness is negligible compared to the shortest observable diffusion length.
In the limit when both the diffusion coefficient $D_m$ of the membrane material and its thickness $l_m$ vanish, the ratio $\kappa \equiv D_m/l_m$ is by definition the {\it permeability}.
The effect of a membrane is described by the boundary condition\cite{Friedman,powles,sukstanskii-jmr2004}
\be \label{BC-gen}
-{\bf n}{\bf J}|_{\r_m} = D_0 {\bf n}\partial_\r \psi|_{\r=\r_m}
= \kappa \lb \psi_{\r_m+{\bf n}0} - \psi_{\r_m-{\bf n}0}\rb.
\ee
Here $D_0$ is the unrestricted diffusion coefficient. The condition (\ref{BC-gen}) means that, at each point $\r_m$ of the membrane, the density $\psi$ of random walkers experiences a jump proportional to the component of the current
${\bf J}$ along the normal ${\bf n}$ to the membrane surface.

The permeability has the dimensions of velocity. In what follows, we find it useful to associate an {\it effective thickness}, $2\ell = D_0/\kappa$, with a membrane. This length scale is defined relative to the free diffusion coefficient $D_0$. Its physical meaning is derived from the condition (\ref{BC-gen}) and is illustrated in Fig.~3a; the membrane indeed appears $D_0/D_m \gg 1$ times thicker than its ``nominal" vanishing thickness $l_m$.



Consider now diffusion in a macroscopic sample embedded with multiple randomly placed membranes, each one imposing the condition (\ref{BC-gen}).
The diffusion propagator depends on their number, shape, and spatial distribution. The number of membranes is characterized by the ratio $S/V$ of their total surface area to the sample volume. 
Here we adopt the convention from the porous media literature\cite{Mitra92,latour-pnas}: a membrane has two faces, so that the membrane's surface area is counted twice in $S$.
The shape and the spatial distribution of the membranes vary greatly depending on the physical context. However, as it will follow from our treatment, the main dispersive features of transport can be captured by making the simplest assumption which also allows us to keep the number of parameters to a minimum. Namely, below we consider the membranes as infinite $d-1$ dimensional planes placed and oriented in a completely random (uncorrelated) way, dividing the sample into pores with random shapes, as shown in Fig.~1 for $d=2$. In this case, the ratio $S/V$ is all what is needed to characterize the geometry. 

In what follows, we first outline our main results for the model medium. Next, we derive them, compare with numerical simulations, discuss and generalize.

\begin{figure}
\includegraphics[width=3.5in]{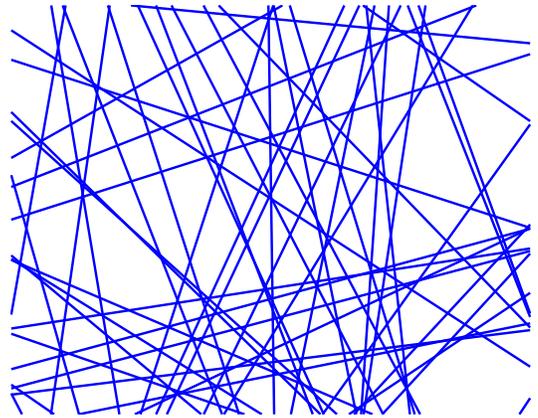}\\
\caption{A fragment of a two-dimensional patch with randomly placed and oriented membranes (one of the disorder realizations used in the simulations).}
\label{fig:randomlines}
\end{figure}

\begin{figure}
\flushleft{\bf a}%
\includegraphics[width=3.0in]{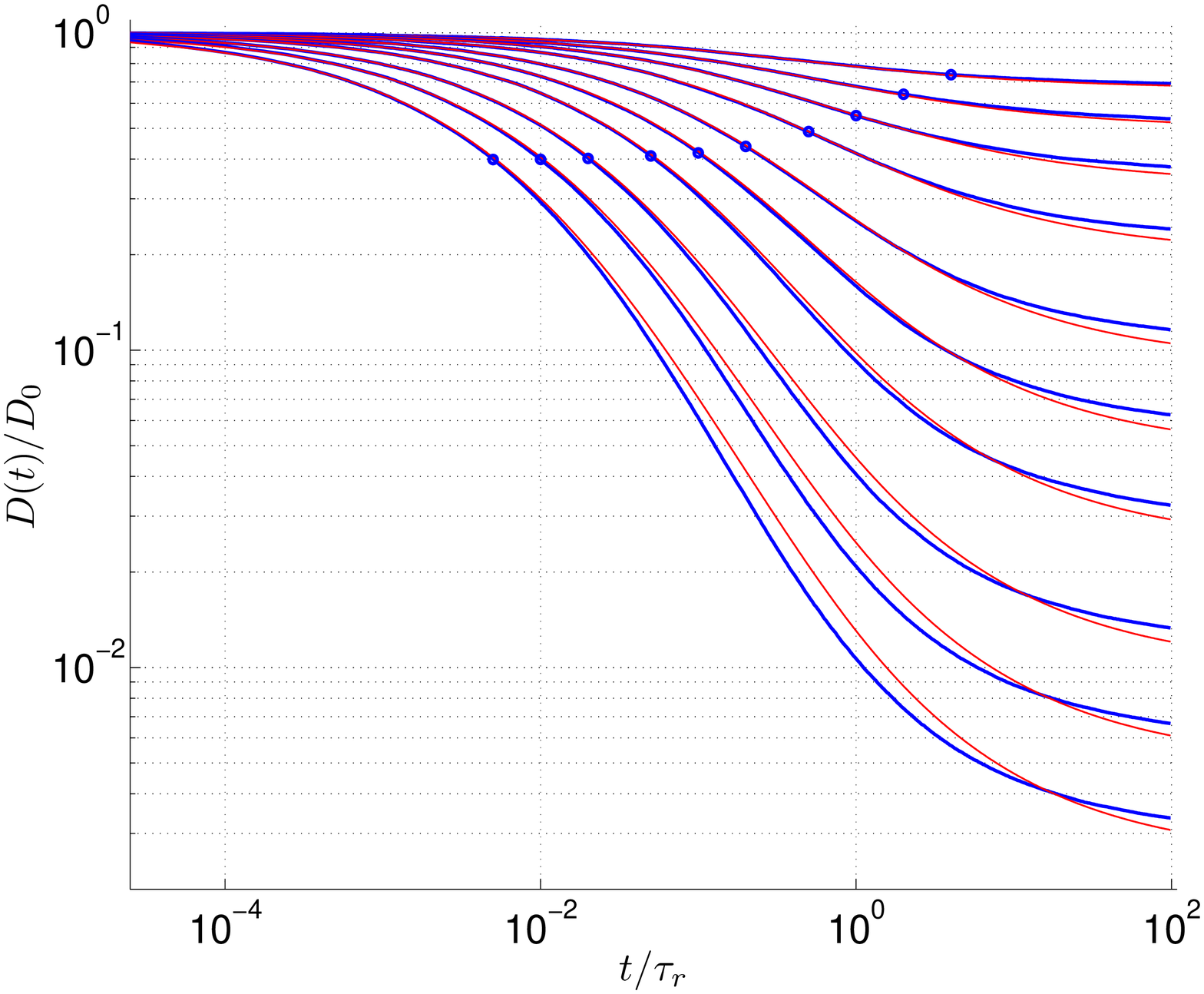}%
\flushleft{\bf b}%
\includegraphics[width=3.0in]{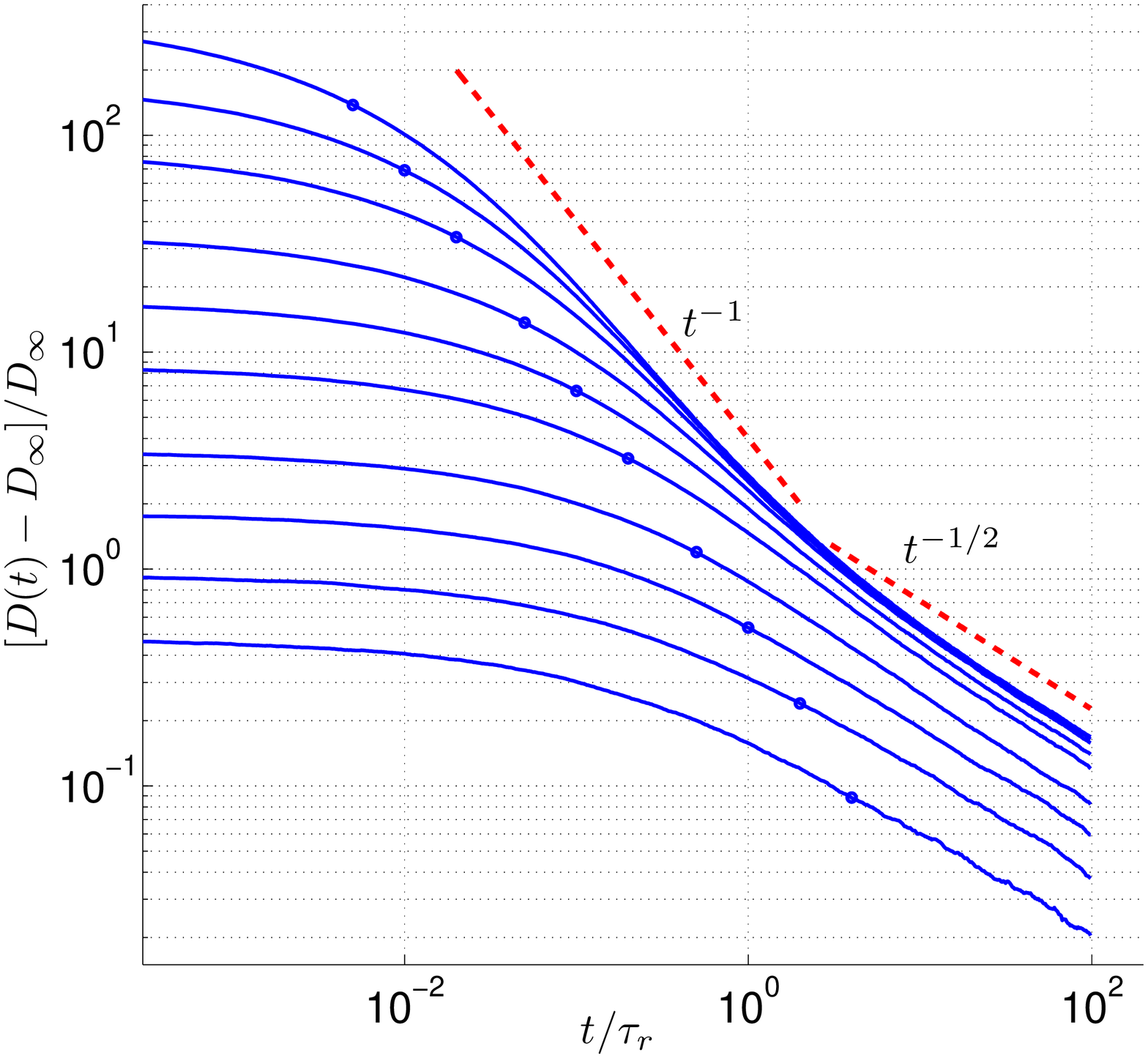}\\
{\bf c}\includegraphics[width=2.5in]{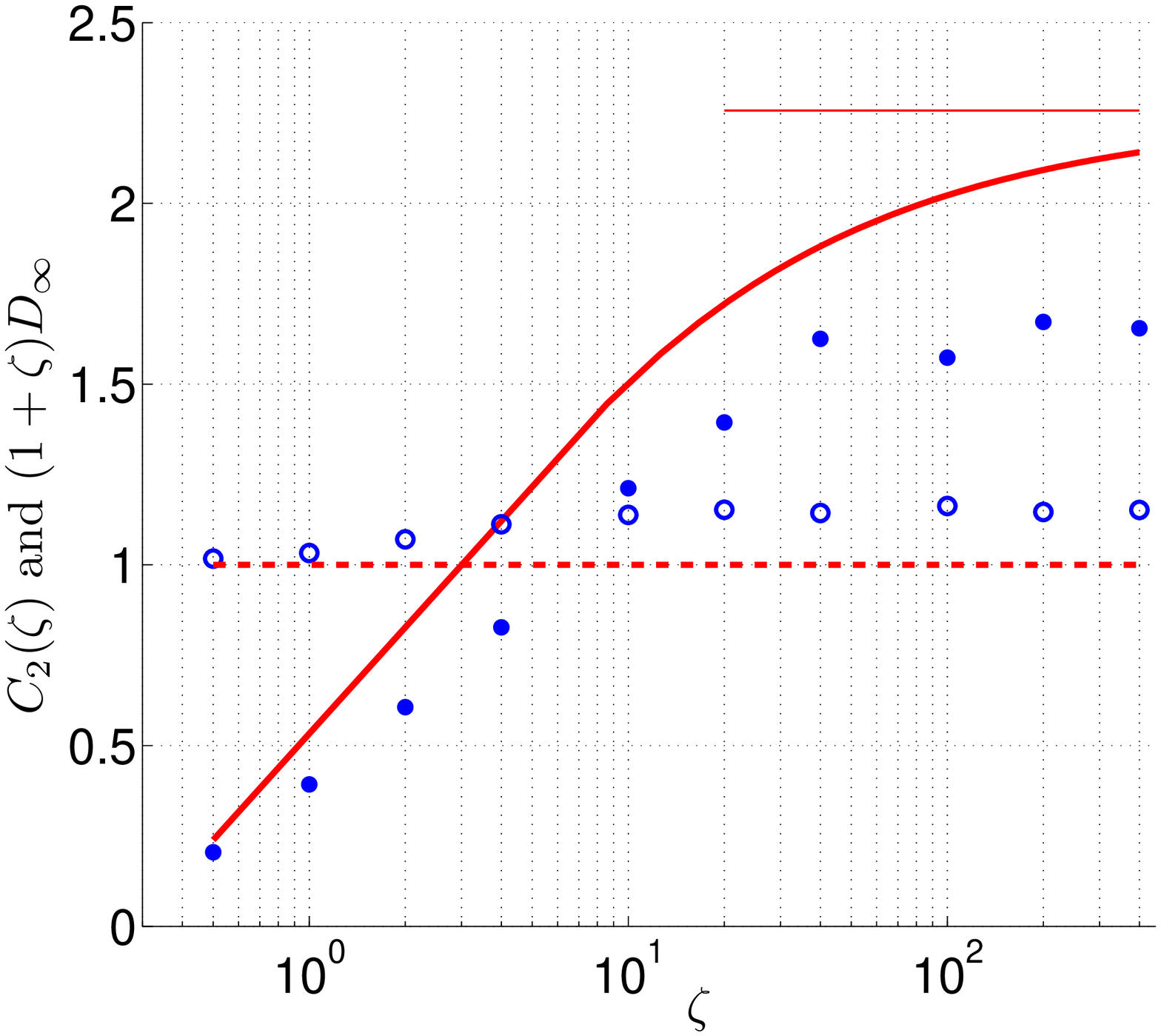}%
\caption{
Time-dependent diffusion coefficient $D(t)$ for the two-dimensional random medium of Fig.~1.
{\bf a}, Comparison of the RG solution (\ref{D-omega-RG}), red, with the Monte Carlo simulations,  blue, for the set of decreasing permeabilities, corresponding to $\zeta=0.5$, 1, 2, 4, 10, 20, 40, 100, 200, 400 (top to bottom). The diffusion time $\tau_D$ is marked by blue circles.
{\bf b}, Scaling behavior of $D(t)$. As the strength $\zeta$ of the restrictions increases from bottom to top, the numerical curves begin to collapse as a signature of the universal behavior (\ref{scaling}).
Dashed lines show the $\zeta\to\infty$ limits from our RG solution:
the ``impermeable" limit $D(t)/D_0 = 2\tau_D/t$
and the scaling limit (\ref{scaling}) with $C_2(\infty)=4/\sqrt{\pi}$.
{\bf c},
Parameters of the scaling limit (\ref{scaling}), $C_2(\zeta)$ (filled circles) and $D_\infty(\zeta)$ (open circles),
determined from the fit of the simulations in {\bf b} to equation (\ref{scaling}), compared with the RG predictions (solid and dashed red lines).
Thin solid line is the RG limit $C_2(\infty)=4/\sqrt{\pi}$.
}
\label{fig:Dt}
\end{figure}

\begin{figure}[b]
\flushleft{\bf a}
\includegraphics[width=3.5in]{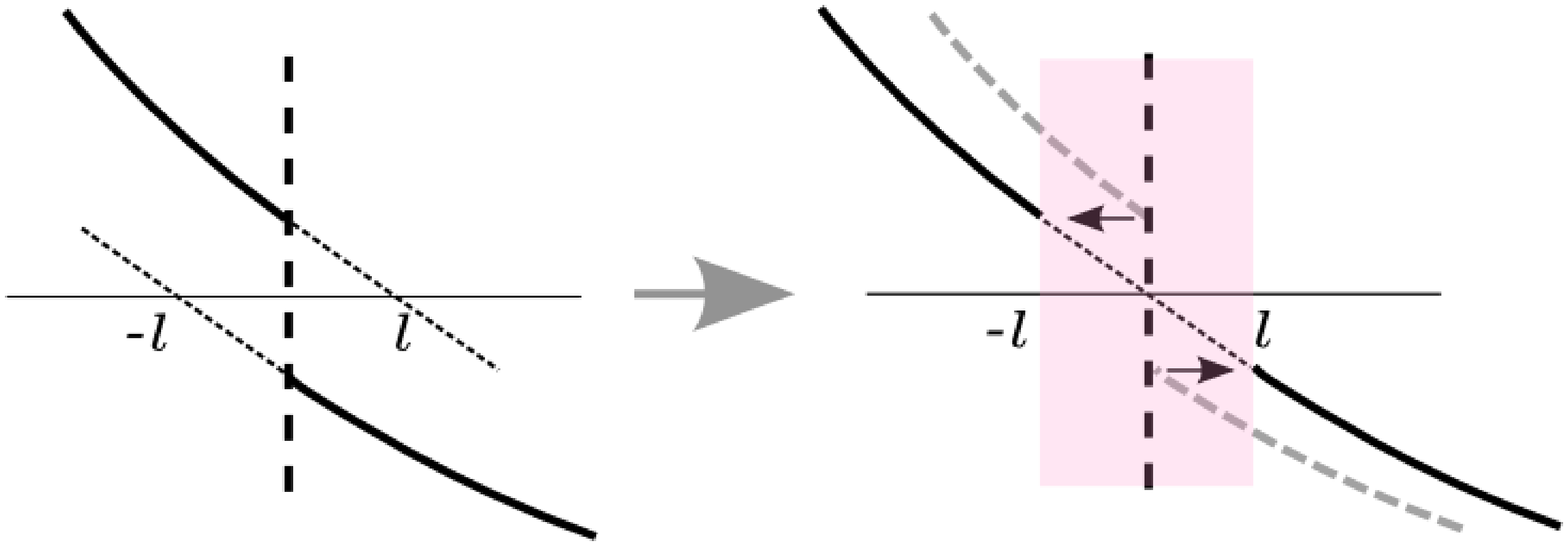}\\
\flushleft{\bf b}
\includegraphics[width=3.5in]{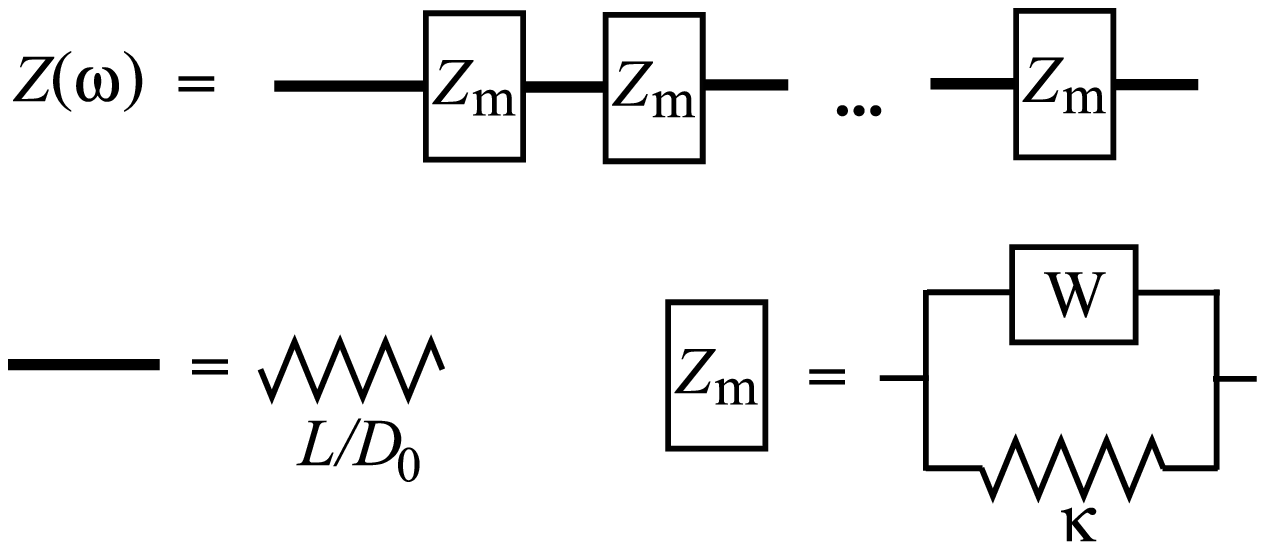}\\
\caption{
{\bf a,}
Meaning of effective membrane thickness: $2\ell=D_0/\kappa$
is the distance by which one
should shift the density profile $\psi(x)$ on each side of membrane if one were to heal the jump discontinuity (\ref{BC-gen}).
{\bf b,}
The equivalent circuit for membranes embedded in a medium with diffusivity $D_0$.
The membrane ``impedance" can be represented as a resistor with conductance $\kappa$ in
parallel with the permeability-independent Warburg element W with conductance $g_W(\w)=\sqrt{i\w D_0}/2i$.
}
\label{fig:def-ell}
\end{figure}

\section*{Results for randomly oriented flat membranes}

In the $t\to\infty$ limit, the diffusion becomes Gaussian with the reduced diffusion coefficient
\be \label{zeta}
D_\infty \simeq {D_0\over 1+\zeta}\,, \quad \zeta = {S\ell \over V d} \,.
\ee
Here the dimensionless parameter $\zeta$ quantifies the ability of membranes to hinder the diffusion. It is the ``volume fraction" occupied by the membranes
based on the effective thickness $\ell$ defined above. The result for $D_\infty$ is exact in $d=1$, ref.~\onlinecite{Crick}. As follows from Fig.~2, it is a good approximation for $d>1$, with $1/d=\la \cos^2\theta\ra$ in equation (\ref{zeta}) arising from the mean-field orientational averaging, cf. Methods section.
Strong restrictions correspond to $\zeta > 1$, when the domains of thickness $\sim \ell$ associated with each membrane overlap, such that transport across a membrane is affected by its neighbors.


For finite $t$, we shift to the frequency representation since the restrictions are stationary. Technically, we focus on the diffusion propagator $G$
averaged over the disorder in positions and orientations of the random membranes. After disorder averaging, the propagator becomes translation invariant.
Its pole in the frequency -- wave vector representation, $G^{-1}_{\w,\q} = -i\w + \D(\w)q^2 + {\cal O}(q^4)$, defines the dispersive diffusion coefficient $\D(\w)$ which is a retarded response function that relates the disorder-averaged particle current
${\bf J}_{\w,\r} = - \D(\w) \partial_\r \psi_{\w,\r}$ to the density gradient.
The corresponding time-dependent diffusion coefficient
\be \label{Dt=Dw}
D(t) \equiv \frac{\la x^2(t)\ra}{2t}= -{1\over t}\,
\int\! {d\omega\over 2\pi}\, e^{-i\w t} {\D(\w) \over (\w+i0)^2}
\ee
is given by a contour integration along the real axis with all the singularities in the lower half-plane of complex $\w$.

We find $\D(\w)$ in three steps. First, we develop a scattering approach for the transmission events described by the boundary conditions (\ref{BC-gen}) at each membrane. Second, we solve the problem perturbatively in the volume fraction $\zeta$ at the mean-field level, starting from infinitely permeable limit $\D(\w)\equiv D_0$. These two steps are done in the Methods section. The dispersive diffusion coefficient, valid up to ${\cal O}(\zeta)$, is
\be \label{D-omega}
\D(\w) = D_0\lp 1 - {\zeta \over 1-z_\w}  \rp
, \quad \zeta\ll 1 \,.
\ee
Here $z_\w = i\sqrt{i\w\tau}$, and $\tau = \ell^2/D_0 = D_0/(2\kappa)^2$.

At the third step we extend equation~(\ref{D-omega}) onto the non-perturbative domain $\zeta\gg 1$ using the real-space renormalization group to account for the multiple transmission events self-consistently, as described below. Thereby we obtain our main result for all $\zeta$,
\be \label{D-omega-RG}
{D_0 \over \D(\w)} = 1 + \zeta + 2z_\w(1-z_\w)\lb \sqrt{1 + \zeta/(1-z_\w)^2} - 1\rb .
\ee

\section*{Effective circuit}

Before deriving our main result (\ref{D-omega-RG}), let us note that
the perturbative limit (\ref{D-omega}) can be represented as a simple effective circuit. As the random walkers are uncharged, the current ${\bf J}$ has only the diffusion component, and the dispersive diffusivity $\D(\w)$ defines its response to a density gradient $\partial_x \psi$ rather than to a potential bias. With this important distinction, consider interpreting
$\D(\w)$ as the analog of the conductivity, in the spirit of the Einstein relation, albeit at finite frequencies\cite{Scher73}.
This way, the ${\cal O}(\zeta)$ result \eqref{D-omega} can be viewed as an ``impedance" $Z(\w)$ of a hypercube $V=L^d$,
$L^{d-2} Z(\w) \equiv {1/\D(\w)} \approx {1/D_0} + n Z_m(\w)$,
which acts as a one-dimensional disordered transmission line of length $L$ with point impedances
\be \label{Zm}
Z_m(\w) = 1/g_m(\w) \,, \quad g_m(\w) = \kappa\big(1-z_\w\big) \,,
\ee
placed at random positions in series, with density $n=S/(2Vd)$. This addition of independent impedances, a consequence of the mean-field description (see Methods section), will be justified later, when we discuss the origin of the $\sqrt{\w}$ dispersion.

The equivalent circuit for a single membrane \eqref{Zm} is shown in Fig.~3b.
In the dc limit, a membrane acts as a resistor with the ``conductance" $\kappa \equiv g_m(0)$; as $\zeta=2\ell n$, each membrane effectively adds the length $2\ell$ to the original clean wire $L$ if its dc resistance were to match $Z(0)\propto 1/D_\infty$.
At finite frequencies the resistor is shunted by the Warburg element\cite{Warburg} with the ``conductance" $g_W(\w)=-\kappa z_\w = \sqrt{i\w D_0}/2i$  independent of the permeability.
The Warburg impedance, first observed at a flat metal electrode in an electrolyte
by Kohlrausch\cite{Kohlrausch} and Wien\cite{Wien} in the 19th century, is associated with the diffusion-limited response\cite{Warburg}.
Incidentally, the impedance $Z(\w)$ is of the Cole form\cite{Cole'34,Cole'41} with power law exponent $\frac12$.

\section*{Renormalization group}

To calculate the response $\D(\w)$ of the disordered transmission line for finite
$\zeta$, we employ the following scaling argument, which we first develop for $d=1$.
Consider a slab of length $L$
with the diffusivity $\D(\w)|_{L}$. Let us extend the slab to the length $L'=bL$, $b>1$, and rescale; now the slab's length $L'$ in new units is back to $L$, while the original, shorter slab has length $L\to L/b$.
The rescaled slab conductance decreases due to adding
extra $N'-N=N\delta b$ membranes, $\delta b=b-1$.
The diffusivity is then reduced as
\be \label{D-delta-n}
\D(\w)|_{L'} = \D(\w)|_{L}
\lb
1 - \delta n \, { \D(\w)|_{L} \over \kappa - \frac{i}2\sqrt{i\w\D(\w)|_{L}}}
\rb
\ee
according to equation~\eqref{D-omega}, as long as the added membrane density
$\delta n=(N/L)\delta b$ is small, $\delta b\ll 1$.
Since we assumed from the beginning that the membrane positions are uncorrelated, adding a small number of membranes in an uncorrelated way at each step is consistent. Choosing an infinitesimal ${\rm d} n \propto {\rm d} b$, we represent
equation~\eqref{D-delta-n} in the differential form,
obtaining the real-space renormalization group (RG) equation
\be \label{RG}
{{\rm d}\D(\w) \over {\rm d}n}
= -{ \D^2(\w)\over \kappa - \frac{i}2 \sqrt{i\w\D(\w)}} \,.
\ee
This is a telegraph equation for a disordered transmission line.
Alternatively, it can be obtained by adding membranes
in small increments $\delta n$ in a macroscopic sample of a fixed length with bare diffusivity $D_0$, and
applying the relation \eqref{D-omega} treating all previously added membranes in
the effective-medium fashion at each RG step.
Either way, the diffusivity flows to lower values as long as the added membranes have finite permeability $\kappa <\infty$.
In $d$ dimensions, rescaling a hypercube $L^d$ as in ref.~\onlinecite{gang4} is equivalent to rescaling a $d=1$ slab in which the effective one-dimensional membrane concentration $n =S/(2Vd)$. At the mean-field level, the problem is always one-dimensional, with the fraction $1/d$ of membranes exerting full resistance.

Integrating equation~\eqref{RG} yields our main result (\ref{D-omega-RG}).
In the dc limit $\D(0)\equiv D_\infty$, the RG flow $1/D_\infty|_{L'}=1/D_\infty|_{L}+(N'-N)/(\kappa L d)$ recovers the above exact result for $d=1$, if one identifies $L$ with the microscopic scale such that $D_\infty|_{L}\equiv D_0$ (the bare diffusivity) and $(N'-N)/Ld \to n$, the final membrane density. Written as RG flow, however, the evolution of the diffusivity is not tied to a particular microscopic cutoff; indeed, the diffusivity $D_\infty|_{L}$ can itself originate from ``more microscopic" membranes on the scale finer than $L$, and so forth.

We also note an interesting equivalent way of looking at the RG flow: Formally identifying the particle number density $\psi$ entering the diffusion equation with an electrostatic potential, the dc limit corresponds to the electrostatic screening of the one-dimensional
electric field (the slope $\partial_x \psi$) by the dipoles (membranes acting as double layers). Coarse-graining corresponds to more dipoles contributing to the screening \cite{KT}, which renormalizes the effective dielectric constant, $1 \to 1+\zeta$.


\section*{Regimes for $D(t)$ and numerical simulations}



Here we discuss the time dependence (\ref{Dt=Dw}) of our result (\ref{D-omega-RG}) and compare it in Fig.~2 with Monte Carlo simulations of diffusion in the two-dimensional geometry of Fig.~1.

First we note that, for $\zeta\sim 1$, the agreement between theory and simulations is very good, which illustrates that the RG framework is able to extend the perturbative result (\ref{D-omega}) to the case of moderate disorder.
Next, we turn to the more interesting and general case of the strong disorder, $\zeta\gg 1$, where there are three distinct regimes in $D(t)$, separated by the two time scales, $\tau_r \gg \tau_D$, defined below.

The initial decrease of the diffusion coefficient
\be \label{Dshort}
D(t)\simeq D_0\lb 1 - {S\over Vd} \lp {4\sqrt{D_0t}\over 3\sqrt{\pi}} -\kappa t \rp\rb
\ee
reproduces the well-known universal short-time $\sqrt{t}$-expansion \cite{Mitra92}.
The correction to free diffusion
arises from the random walkers within the diffusion
length $\sim \sqrt{D_0 t}$ from the barriers; the term $\propto t$ is the permeability effect \cite{sen-perm} for the case of flat pore walls.
The limit (\ref{Dshort}) is valid as long as only a small fraction of random walkers has encountered the pore walls, i.e. for $t\ll \tau_D$, where
$\tau_D = \bar a^2/2D_0$ is the diffusion time across the typical pore size
$\bar a \simeq 1/n=2d V/S$. Equation~(\ref{Dshort}) follows from the first two terms of the $|z_\w|\gg \zeta$ expansion of (\ref{D-omega-RG}) and subsequent application of (\ref{Dt=Dw}).
This expansion can be performed using the perturbative limit (\ref{D-omega}), as
it becomes universally valid for large frequencies, $\w\tau_D\gg 1$, when only the walkers within the wavelength $\sim \sqrt{D_0/\w}$ from the membranes are restricted. There is a perfect agreement between theory and simulations in this regime for all $\zeta$.


Highly restrictive membranes look completely impermeable for $t\ll \tau_r$, where $\tau_r=V/(\kappa S)$ is the residence time\cite{Friedman} within a typical pore; in our notation, $\tau_r = (\zeta/d)\tau_D$.
As a result, the mean square displacement $\la x^2\ra$ is bounded by
$\sim \bar a^2$, and $D(t)/D_0 \sim \tau_D/t$.
For our RG result (\ref{D-omega-RG}), the ``impermeable" behavior occurs for
$1\ll \sqrt{\zeta}\ll |z_\w| \ll \zeta$, and yields
the purely imaginary $\zeta$-independent limit $\D(\w)\simeq -i\w \bar a^2$ (similar to the purely reactive electrical conductivity),
corresponding to $D(t)/D_0 = 2 \tau_D/t$, Fig.~2b.
Both our result (\ref{D-omega-RG}) and the simulations display the $1/t$ dependence for $\zeta\gtrsim 100$, with the theory somewhat overestimating the $D(t)$ obtained from the numerics for very large $\zeta$. The discrepancy between the RG result and the simulations is maximal for $t\sim \tau_r$ as expected from the discussion of our effective medium approach (see the Methods section).

For times longer than $\tau_r$, the system becomes aware of the finite leakage across the membranes, with $D(t)$ approaching the limit (\ref{zeta}) which can be qualitatively estimated as hopping on a $d$-dimensional lattice with a step $\bar a$ over the time $\tau_r$, $D_\infty \sim \bar a^2/(\tau_r d)$.

Remarkably, the way $D(t)$ approaches $D_\infty$ slows down from $\sim 1/t$ to  $1/\sqrt{t}$-decrease in any dimensionality $d$, defining the novel disorder universality class, represented by the randomly placed infinite membranes:
\be \label{scaling}
D(t) \simeq
D_\infty \lp 1 + C_d(\zeta) \, \sqrt{\tau_r \over t}\ \rp ,
\quad t \gtrsim \tau_r \,.
\ee
This follows from $\D(\w)=D_\infty + {\cal O}(\sqrt{\w})$ in the limit $|z_\w|\ll \sqrt{\zeta}$. The regime (\ref{scaling})
can be clearly seen from the simulation results of Fig.~2b
where time is in the units of $\tau_r$ for a set of permeabilities.
In this limit, the relative deviation $[D(t)-D_\infty]/D_\infty \lesssim 1$.
We emphasize that for $\zeta \to\infty$, the dependence (\ref{scaling}) approaches a universal scaling law with a fixed $C_d=C_d(\infty)$, which can be represented only in terms of the effective parameters $D_\infty$ and $\tau_r$, independent of the original microscopic parameters $D_0$, $\kappa$ and $S/V$. Approaching this law corresponds to the collapse of the simulation curves in Fig.~2b onto one universal curve for $\zeta\gtrsim 100$ and $t\gtrsim \tau_r$. Our RG approximation, yielding $C_d(\infty) = \sqrt{8d/\pi}$, overestimates the $C_2(\infty)$ obtained from the fit of the numerical curves to equation (\ref{scaling}), as illustrated in Fig.~2c.
For finite $\zeta$, the parameters $D_\infty$ given by equation (\ref{zeta}) and
$C_d(\zeta)=C_d(\infty)\sqrt{\zeta}\lp \sqrt{1+\zeta}-1\rp/(1+\zeta)$
agree with the simulations within about 15\% and 30\% correspondingly, even in the strongly non-perturbative regime of $\zeta \gtrsim 100$.

The fact that the RG solution works fairly well even when the ``small parameter" $\zeta\gg 1$ can be understood in terms of the flow to the moderately disordered limit with strongly renormalized parameters, $D_0\to D_\infty$ and $\tau\to \tau_r$.
Indeed, for $t\gg \tau_r$, the medium effectively looks as if it had a decreased diffusion coefficient $D_\infty$. This, in turn, {\it reduces the contrast} between the diffusion coefficient inside the membrane and of its surroundings, effectively changing $D_0\to D_\infty$ in the boundary condition (\ref{BC-gen}), thereby reducing its effective thickness $\ell \to \tilde \ell = D_\infty/(2\kappa) \simeq \ell/\zeta \sim \bar a$ down to the mean distance between membranes. Hence, the renormalized disorder strength is reduced,
$\zeta \to \tilde \zeta \equiv S\tilde \ell /Vd \sim 1$.
Likewise, the renormalized time scale
$\tau \to \tilde\tau \equiv \tilde \ell^2/D_0\sim \tau_r$ matches the residence time.
At this point the perturbative limit (\ref{D-omega}) is matched.

Conversely, when the membranes are very permeable, $\zeta\lesssim1$, the residence time in a pore is of the order of $\tau_D$, and the time scale $\tau_r$ defined above becomes obsolete. In the perturbative limit $\zeta\ll 1$ our result (\ref{D-omega}) becomes exact,
the intermediate $1/t$ regime does not appear, and the short-time behavior (\ref{Dshort}) directly crosses over to the nonanalytic dispersion of the form (\ref{scaling}),
\be \label{Dpert-asy}
D(t)|_{t\gg \tau} \simeq D_\infty\lp 1 + {2\zeta\over \sqrt{\pi}} \sqrt{\tau\over t} \rp,
\quad \zeta\ll1 \,.
\ee
The role of $\tau_r$ is now played by the time scale $\tau = \zeta (d/2) \tau_r$
defined after equation~(\ref{D-omega}).
Hence it is now the time scale $\tau\ll \tau_r$ which is associated only with the membrane properties and not with their density $S/V$, that determines the dispersion of diffusion for randomly placed highly permeable membranes, as discussed below.

\section*{Origin of the $\sqrt{\w}\sim 1/\sqrt{t}$ dispersion}


The defining feature of our model, the non-analytic dispersion $\D(\w)-D_\infty \sim \sqrt{\w}$, is present already at the level of the perturbative result (\ref{D-omega}). We now discuss its physical origins, as well as justify
the mean-field approach utilized to obtain equation (\ref{D-omega}) and its nonperturbative counterpart (\ref{D-omega-RG}).

In the perturbative $\zeta\ll 1$ limit, the two relevant time scales are $\tau_D$ and $\tau = \frac12 \zeta^2 \tau_D$. We first focus on the limit $t\ll \tau_D$, where adding independent contributions  (\ref{Zm}), cf. equation (\ref{sigma1}), is valid since multiple scatterings off different membranes are not yet important. In this case, the dispersive behavior (\ref{Dpert-asy}) for $\tau \ll t \ll \tau_D$ is a result of scattering off individual membranes, in the following sense. Consider a constant density bias $\Psi_0$ introduced across a membrane at $t=0$. The single-membrane conductance (\ref{Zm}) yields the transient current response $J(t)=\kappa \Psi_0\big(1-P_{\rm surv}(t)\big)$ which involves the survival probability $P_{\rm surv}|_{t\gg \tau} = \sqrt{\tau/\pi t}$ of a random walker released a distance $\ell = \sqrt{D_0\tau}$ from the origin, to never return to the origin during time $t$.
The fraction $P_{\rm surv}(t)$ represents the walkers who have managed to wander around the ``biased" side and have not yet encountered the membrane.
Hence, the transmission process is represented in terms of multiple attempts to return to the position of the membrane with net probability $1-P_{\rm surv}(t)$, times the rate $\propto \kappa$ to overcome the restriction.
Equivalently, the conductance kernel
$g_m(t)=\int {d\w\over 2\pi}\, e^{-i\w t} g_m(\w) \equiv \kappa p(t)$ is the L\'evy-flight probability distribution function $p(t)=-\dot P_{\rm surv} \simeq c\, t^{-3/2}$ for time intervals $t$ between successive returns of a
Brownian particle to the origin \cite{Bouchaud}, with
$\tau$ being the short-time cutoff, and $c=\sqrt{\tau/4\pi}$.
Note that the contribution $\kappa P_{\rm surv}(t)$ of the ``survivors", originating from the Warburg contribution $g_W$ to (\ref{Zm}), is independent of the permeability, as they have not yet encountered the membrane.
In the dc limit $t=\infty$ 
the membrane acts as a simple resistor since in one dimension all Brownian particles eventually return to the origin. Incomplete transmission for $\tau\ll t \ll \tau_D$ produces the $1/\sqrt{t}$ correction to the dc ``conductance".


For longer times $t\gg \tau_D$, multi-membrane scatterings determine the time dependence of $D(t)$ which becomes sensitive to correlations between positions of membranes in space. Indeed, at large diffusion times, the medium is effectively coarse-grained into domains of the size
$\sim L(t)=\sqrt{2D t}\gg \bar a$. These domains have slightly different diffusion coefficients $D_j$ due to a different number $N_j$ of membranes falling into them.
The diffusion coefficient $D(t)\equiv \langle D_j\rangle$ is the ensemble average over the domains; in one dimension, its relation to $D_\infty$ is determined by adding the ``dc resistances", $1/D_\infty = \langle 1/D_j\rangle$, leading to $D(t) \simeq D_\infty + \langle (\delta D)^2\rangle/D_\infty$.
The variance $\langle (\delta D)^2\rangle \sim \langle (\delta N)^2\rangle/L^2\sim 1/L(t)$
scales as $t^{-1/2}$ as a result of the Poissonian statistics in the membranes' positions.
This simple argument elucidates the meaning of the previously obtained one-dimensional results\cite{Machta81,Zwanzig,Denteneer-Ernst,Visscher}. 
(The crucial role of the Poissonian statistics is obvious from a qualitatively different $1/t$ decrease of $D(t)$ in the case of periodic membranes in $d=1$.\cite{sukstanskii-jmr2004})

Remarkably, our results (\ref{D-omega}) and (\ref{D-omega-RG}) are compatible with the statistics of restrictions provided by randomly placed membranes in {\it all} dimensions $d$. The scaling $\langle (\delta D)^2\rangle \sim 1/L(t)$ persists for $d>1$
as the random infinitely long membranes generate qualitatively stronger fluctuations in the distribution of $\{ D_j\}$, exemplifying the role of the spatially correlated disorder introduced here.
Furthermore, the prefactors of the $1/\sqrt{t}\sim \sqrt{\w}$ terms of equations (\ref{Dpert-asy}) and (\ref{D-omega}) appear to be exact in $d=1$, as it can be checked by generalizing the lattice calculation of ref.~\onlinecite{Denteneer-Ernst} onto our case of the membranes having random positions on a line, validating our results for $t\gg \tau_D$.
This is why the numerical simulations for small $\zeta$ agree with theory very well even for very long $t$ (Fig.~2). This observation allows us to conclude that adding independent membrane contributions in a mean-field way is compatible with the Poissonian statistics of the disorder for all $\w$. Formally this allows us to extend the above connection with the return-to-origin probability onto $t\gg \tau_D$ by substituting the rest of the system via a ``featureless" effective medium with diffusivity $D_\infty$.
The equivalence between Poissonian disorder in membranes' positions and the return-to-origin probability persists in the RG solution, as it is utilized at every RG step (\ref{D-delta-n}). As a result, these two pictures remain equivalent in the $\zeta\gg 1$ limit after the effective-medium substitution $\tau\to\tau_r$, and the scaling behavior (\ref{scaling}) is justified for all $\zeta$ and $t$.



\section*{Outlook}

In this work we have introduced a novel class of disorder, represented by straight permeable membranes which are randomly placed and oriented. A random medium of this type has dispersive diffusion $\D(\w)-D_\infty \sim \w^{1/2}$ in any dimensionality $d$. This dispersion for $d>1$ is notably more singular than its well studied counterparts giving $\w^{d/2}$ in spatially uncorrelated 
random media\cite{Machta81,Zwanzig,Denteneer-Ernst,Visscher,Ernst-I}.
As the $\w^{1/2}$ singularity originates from long-range spatial correlations introduced by the membranes, the dispersive behavior of the form (\ref{scaling}) will persist in a variety of samples with random locally flat restrictions, and will become increasingly important for longer diffusion times when the contributions from shorter-ranged disorder types vanish.
However, the behavior (\ref{scaling}) will eventually terminate when the membranes cease to be flat past their intrinsic correlation (gyration) radius $r_c$, i.e. for $t > t_c\sim r_c^2/D(t_c)$, when the structural memory is forgotten, and the dimensionality-specific dispersion $\sim \w^{d/2}$ of a coarse-grained random medium sets in.
\section*{Methods}

{\bf A single membrane at the origin.}
Consider first a one-dimensional problem.
The boundary condition (\ref{BC-gen}) can be represented as a local scattering term
in the diffusion equation
\be \label{DE}
\partial_t \psi = D_0\partial_x^2 \psi + \V_x \psi \,, \quad
\V_x \psi \equiv -2D_0\ell \delta'(x) \lb \partial_x \psi\rb_{x=0} \,.
\ee
Here $\delta'(x)$ is the derivative of the Dirac delta-function at the position
of the membrane.

We focus on the Green's function $\G_{t;x,x'}$ of the problem
\eqref{DE}, defined by the initial condition $\G_{t;x,x'}|_{t=0} = \delta(x-x')$.
As the problem is stationary, we shift to the frequency
domain, $\partial_t \to -i\w$. In this representation,
the Green's function $\G_{\w; x,x'}$ is formally an operator inverse
\be \label{G=inv}
\G_{\w} = {\lb {\Gbare_\w}^{-1} - \V \rb^{-1}}
\ee
where the coordinates $x$ play a role of indices in infinite-dimensional matrices,
and the free propagator
\be \label{Gbare}
\Gbare_{\w,q} = {1\over -i\w + D_0 q^2} \,.
\ee
The result of inversion \eqref{G=inv} is the Born series
\be \label{G-born}
\G_{\w;x,x'} = \Gbare_{\w;x-x'}
+ \int\! {\rm d}x_1\, \Gbare_{\w; x-x_1} {\bf V}_{\w;x_1} \Gbare_{\w;x_1-x'}
\ee
where the key quantity, the so-called ``full vertex''
satisfies the Dyson equation, which in the Fourier representation reads
\be \label{dyson-k}
{\bf V}_{\w;k,k'} = V_{k,k'}
+ \int\!{{\rm d}q\over 2\pi}\, V_{k,q}\Gbare_{\w,q} {\bf V}_{\w;q,k'} \,.
\ee
In our case, the bare vertex is separable,
\be \label{Vkk'}
\V \to V_{k,k'} = 2D_0 \ell \, kk' \,.
\ee
The solution of equation~\eqref{dyson-k}, the full vertex
\be \label{Vfull}
{\bf V}_{\w;k,k'} = {2D_0\ell \, kk' \over 1 -i\sqrt{i\w\tau}}\,.
\ee
Throughout this work, $\w$ is understood as having an infinitesimal positive imaginary part $+i0$ due to causality; this way the retarded response functions are analytic in the upper-half-plane of the complex variable $\w$.

The $\w^{1/2}$ singularity parallels that of the energy-dependent quantum scattering off a localized potential in one dimension, identifying $i\w \to \epsilon+i0$. The difference, however, is that in the quantum problem, energy comes with a negative power $\epsilon^{-1/2}$ in the denominator of the full vertex: Scattering is maximum for high-energy particles, whereas low-energy ones do not notice a localized potential. In our case the situation is opposite: A membrane is most important in the dc limit, with ${\bf V}_{\w;k,k'}\to V_{k,k'}$, whereas at short times its scope is reduced down to the particles within the diffusion length $\sim \sqrt{D_0/\w}$, with ${\bf V}_{\w;k,k'}\sim V_{k,k'}/\sqrt{\w}$.

{\bf Multiple membranes.}
Averaging over positions of randomly placed membranes with concentration $n$
is done in a standard way \cite{haus-kehr,Bouchaud,Altshuler-Aronov},
by introducing the self-energy part $\Sigma_{\w,q}$,
\be
\label{G-sigma}
G^{-1}_{\w,q} = {\Gbare_{\w,q}}^{-1} - \Sigma_{\w,q} \,.
\ee
On the mean-field level, we keep in $\Sigma_{\w,q}$ only the exact interaction
vertex with a single membrane,
\be \label{sigma1}
\Sigma_{\w,q} \simeq n {\bf V}_{\w;q,q} \,.
\ee
The mean-field approximation is correct in the first order in membrane
concentration $n = S/2V$ for all frequencies, as justified in section on the origin of the $\sqrt{\w}$ dispersion.

The self-energy shifts the pole of the propagator $G_{\w,q}$,
thereby changing the diffusive dynamics due to disorder-averaged interaction
with random membranes. Remarkably, on this level there are no
higher-order terms in powers of $q^2$.
This allows us to represent interaction with membranes solely as
renormalization of the effective diffusivity $D_0\to \D(\w)$, equation~\eqref{D-omega}, which acquires frequency dependence dictated by that of the vertex \eqref{Vfull}.
The role of $\D(\w)$ is similar to that
played by the dispersive refraction index. Indeed,
the relation (\ref{sigma1}) is analogous to the mean-field relation
between the refraction index and the forward scattering amplitude\cite{Landau3}.
In our case, 
the presence of ``scatterers'' (membranes) introduces the memory kernel into the current-density response.

In higher dimensions $d>1$, averaging is performed both over positions of membrane
in the direction $\n$ normal to its surface, and over its orientations.
This amounts to a frequency shift $i\w \to i\w(\q)=i\w -D_0\q_\parallel^2$,
$\Sigma_{\w,\q;\n} = (S/2V){\bf V}_{\w(\q);\q\n, \q\n}$,
with $\q_{\parallel}=\q-(\q\n)\n$ the conserved momentum parallel to membrane,
and subsequent orientational average
$\Sigma_{\w,\q} \equiv \la \Sigma_{\w,\q;\n}\ra_\n$.
Substituting the latter self-energy part in equation~\eqref{G-sigma},
obtain the result \eqref{D-omega} for any $d$.
Note that the $\sqrt{\w}$ dispersion,
a signature of the one-dimensional character of the scattering problem in
the direction normal to the membrane, survives the orientational average in $d>1$.

{\bf Effective medium picture.}
The RG approach assumes that the boundary condition \eqref{BC-gen} is adjusted accordingly, $D_0\to \D(\w)|_L$ at each RG scale $L$.
Strictly speaking, this is valid only for sufficiently small frequencies $\sqrt{\D(\w)/\w} \gg 1/n$, corresponding to diffusing past many membranes, when the system can be treated as a uniform effective medium at each RG step. Luckily, for the opposite case of large frequencies the full result \eqref{D-omega-RG} matches the exact $\sqrt{\w\tau}\gg \zeta$ limit \eqref{D-omega}, corresponding to the original boundary condition \eqref{BC-gen}. This allows us to keep the frequency simply as a parameter in equation~\eqref{RG}.

From this discussion one expects that the agreement of the RG result (\ref{D-omega-RG}) with the simulations
is the worst in the intermediate regime, when $\sqrt{\D /\w} \sim 1/n$,
equivalent to $t\sim \tau_r$. Fig.~2a shows that the discrepancy between the RG result and the simulations is indeed highest around $t\sim \tau_r$ and remains under 25\%.
The agreement improves for $t\gg \tau_r$ to within 10\% even for the largest $\zeta$.

We note that in $d=1$, the RG limit $C_1(\infty)$ overestimates the exact result for the one-dimensional lattice model\cite{Machta81,Zwanzig,Denteneer-Ernst,Visscher}, $\sqrt{2/\pi}$, by the factor of two. We also observe this in the one-dimensional simulations (not shown). There are no exact results for this problem for $d>1$. As it is often true with mean-field-like description, the accuracy of our solution is likely to improve for increasing $d$. This is consistent with the notably better agreement (Fig.~2c) of the theoretically obtained coefficient $C_2(\zeta)$ with that determined from the $d=2$ simulations.

{\bf Monte Carlo dynamics} was realized on a square patch with periodic boundary conditions, embedded with randomly placed and oriented membranes, cf. Fig.~1, of nearly identical $S/V$. For the simulation of each $\zeta$ in Fig.~2, the diffusion coefficient $D(t)$ was calculated by averaging the displacement variance $\la x^2\ra$, cf. equation (\ref{Dt=Dw}), over a total of $4\times 10^5$ random walkers evenly split between 40 disorder realizations, with an average of 385 membranes per patch. The trajectory of each random walker was a sequence of moves in a randomly chosen direction over a distance $\d r=\sqrt{4D_0\d t}$ during a time step $\d t$, with the total diffusion time $t$ up to $100\tau_r$, corresponding to a maximum of $4\times 10^6$ time steps per walker.
Transmission across a membrane occurred with probability $P\propto \kappa \d r/D_0 \ll 1$, cf. ref.~\onlinecite{powles}. The disorder strength $\zeta$ was varied by changing $\kappa$.
The time step $\d t$ for each $\zeta$ is chosen so that $P<0.007$, and the ratio $\d r/\bar a = S\d r/4V < 0.1$. We calibrated our results for a quasi-one-dimensional disorder (random parallel membranes), which reproduced the exact limit (\ref{zeta}) with about 1\% accuracy.
The random walk simulator was developed in C++. Simulations were performed on the NYU General Cluster. With an average of 200 CPU cores used simultaneously, all simulations took about 100 hours.


\begin{acknowledgments}
We thank Valerij Kiselev and Daniel Sodickson for discussions.
Research was supported by the Litwin Fund for Alzheimer's Research, and the
National Institutes of Health Grant 1R01AG027852.
\end{acknowledgments}


\end{document}